\documentclass[11pt,onecolumn, draftclsnofoot]{IEEEtran}

\usepackage{amsmath}

\usepackage{graphicx}
\usepackage{amssymb}
\begin{document}
\newtheorem{definition}{Definition}
\newtheorem{condition}{Condition}
\newtheorem{corollary}{Corollary}
\newtheorem{example}{Example}
\newtheorem{lemma}{Lemma}
\newtheorem{theorem}{Theorem}
\newtheorem{proposition}{Proposition}
%
\renewcommand\theenumi{\alph{enumi}}
\renewcommand\theenumii{\roman{enumii}}

\title{Throughput-Delay Analysis of Random Linear Network Coding for Wireless Broadcasting}

\author{\IEEEauthorblockN{B. T. Swapna, Atilla Eryilmaz, and Ness B. Shroff\\Departments of ECE and CSE \\ The Ohio State University \\}
\IEEEauthorblockA{Email: \{buccapat, eryilmaz, shroff\}@ece.osu.edu}}

\maketitle

\begin{abstract}
In an unreliable single-hop broadcast network setting, we investigate the throughput and
decoding-delay performance of random linear network coding
as a function of the coding window size and the network
size. Our model consists of a source transmitting packets of a
single flow to a set of $n$ users over independent
erasure channels. The source performs random linear network
coding (RLNC) over $k$ (coding window size) packets and
broadcasts them to the users. We note that the
broadcast throughput of RLNC must vanish with increasing
$n$, for any fixed $k.$ Hence, in contrast to other works
in the literature, we investigate how the coding window
size $k$ must scale for increasing $n$. Our analysis reveals that the coding window size of $\Theta(\ln(n))$ represents a phase transition rate, below which the throughput converges to zero, and above which it converges to the broadcast capacity. Further, we characterize the asymptotic distribution of decoding delay and provide approximate expressions for the mean and variance of decoding delay for the scaling regime of $k=\omega(\ln(n)).$
These asymptotic expressions reveal the impact of channel correlations on the throughput and delay performance of RLNC. We also show how our analysis can be extended to other rateless block coding schemes such as the LT codes. Finally, we comment on the extension of our results to the cases of dependent channels across users and asymmetric channel model.
\end{abstract}


\section{Introduction}
We consider an important transmission scenario, occurring
in many communication systems, whereby a source must
broadcast common information to many users over
wireless channels in a timely manner. Such a scenario
occurs, for example, in a satellite or cellular network
where a satellite or base station broadcasts a large file
or streaming multi-media data to many users within
their footprint over unreliable channels. Another example
occurs in a multi-hop wireless network where each node
broadcasts control information to all its immediate
neighbors to coordinate medium access, power control, and
routing operations. We note that such local sharing of
control information (such as queue-length or other pricing
information) is common to many provably efficient network
controllers (e.g. \cite{joo1,joo2,linsrikant} etc.).

In this work, the essential components of such wireless
broadcast systems are modeled through a transmitter
broadcasting consecutive blocks of $k$ data packets to $n$ users over
independent and identically fading time-correlated erasure channels with steady state erasure probability $p$. Assuming that the
transmitter is infinitely backlogged, we consider
transmission strategies that transfer the data in blocks of
$k$ packets, which include the class of block coding
strategies. Among all such block transmission strategies,
it has previously been shown (see \cite{Eryil})
that, for any fixed $n$ and $k,$ the Random Linear Network
Coding (RLNC) strategy (see Section~\ref{sec:model} for a
detailed description) asymptotically\footnote{This
asymptotic is with respect to increasing field size over
which the data packets are defined (see
Section~\ref{sec:model}).} minimizes the number of
transmissions required to complete the transfer of all $k$
packets at all $n$ users (also called the \emph{block
decoding delay}).

With this motivation, we focus on the scaling
performance of RLNC as a function of $k$ and $n$ with
respect to the following two key metrics: the
\emph{(broadcast) throughput}, defined as the data transfer
rate to all users; and the \emph{(broadcast) decoding
delay}, defined as the amount of time spent between the
start of a block transmission and its completion (i.e.,
successful decoding) at all the users.

It is not difficult to see that the \emph{(broadcast) capacity} of
such a collection of $n$ erasure channels, for any $n,$ is equal to
$(1-p)$ packets per time slot. Moreover, this maximum limit on the
throughput can be arbitrarily closely achieved by encoding
information into an arbitrarily large block size, $k.$ Yet, this is
not attractive since it leads to a decoding delay that diverges to
infinity. In this work, we address the question of whether RLNC can
achieve throughput arbitrarily close to the capacity while yielding
acceptable decoding delay. The main contributions of this work are:
\vspace{-0.05in}
\begin{itemize}
\item We find that the broadcast throughput of RLNC must vanish for
  any fixed $k$ as $n$ tends to infinity. We expose the cause of this
  behavior through a key example (see Section~\ref{sec:related}),
  which motivates our search of a proper scaling of the block size $k$ with increasing $n.$
\item Using upper and lower bounds on the
throughput and decoding-delay performance of the
broadcast system, we show that a \emph{phase transition} in the
performance of our system occurs at the block
length scaling rate\footnote{We use the standard order
notation: $g(n)=o(f(n))$ implies $\lim_{n\rightarrow \infty}
(g(n)/f(n))=0;$ and $g(n)=\omega(f(n))$ implies $\lim_{n\rightarrow \infty}
(g(n)/f(n))=\infty;$ and $g(n)=\Omega(f(n))$ implies
$\lim_{n\rightarrow \infty} (g(n)/f(n))\geq c$ for some constant
$c;$ and $g(n)=\Theta(f(n))$ implies $c_1 \le \lim_{n\rightarrow \infty}
(g(n)/f(n)) \le c_2$ for some constants $c_1$ and $c_2.$} of $k=\Theta(\ln(n))$ with respect to the
network size $n$. Specifically, we show that if $k$ increases slower
than $\ln(n),$ then the broadcast throughput of RLNC converges to
zero, and if $k$ increases faster than $\ln(n)$, then the broadcast throughput of RLNC converges to the broadcast capacity of $(1-p).$ In Section~\ref{sec:related}, we shall note that the nature of this phase-transition is different from the previously observed phase-transition phenomenon in \cite{Starobinski} due to a key difference in the scaling of $k$ and the metric of focus.
\item We characterize the asymptotic distribution of decoding delay and compute the mean and variance of decoding delay for the scaling regime of $k=\omega(\ln(n))$ using extreme value theory.
\item We provide numerical results to substantiate our findings. Our results verify the phase transition with respect to the scaling rate and, not surprisingly, indicate that the rate of convergence to capacity can be increased by choosing a faster scaling of $k$ with respect to $n$ at the cost of higher decoding delay. Our results also show that the approximate expression for mean decoding delay obtained using extreme value theory is accurate even for small values of $n.$
\end{itemize}

These results collectively imply that RLNC can achieve
throughput-delay tradeoff of $((1-p),\Omega(\ln(n))).$
This is an attractive result as it indicates that as long as the
coding block size scales super-logarithmically (i.e., very slowly)
with the network size, the broadcast capacity is achievable with a
simple policy such as RLNC.

The rest of the paper is organized as follows. In
Section~\ref{sec:related}, we overview some of the relevant
work in this context and provide an example that motivates
this work. After introducing the main system components in
Section~\ref{sec:model}, we provide our throughput and
delay analysis of RLNC for the case of time invariant erasure channels in Section~\ref{sec:corranalysis}. We present some numerical results to substantiate our findings in the Section~\ref{sec:num}. In Section~\ref{sec:disc}, we comment on three important extensions of our results - analysis of other rateless block coding schemes such as the LT codes, dependent channels across users, and asymmetric channel model. Finally, our
conclusions are provided in
Section~\ref{sec:conc}.

\section{Related Work and Motivating Example}
\label{sec:related} \indent Our model is similar to that considered
in \cite{Eryil,Starobinski, Ghaderi}. In \cite{Ghaderi}, the authors quantify the reliability gain of RLNC for a fixed coding window
size and show that RLNC significantly reduces the number of
retransmissions in lossy networks compared to an end-to-end ARQ
scheme. The delay performance gains of RLNC were observed in \cite{Eryil}. They show that, for a fixed coding
window size $k$, the network coding capability can lead to
arbitrarily better delay performance as the system parameters
(number of users) scale when compared to traditional
transmission strategies without coding.\\
\indent Also, in a similar setup as in this paper, it has recently been shown
in \cite{Starobinski} that, for any given coding window size, there exists a phase transition
with respect to decoding delay such that there is a threshold on
the number of transmissions below which the probability that a block
of coded packets can be recovered by all the nodes in the network is
close to zero. On the other hand, if the number of transmissions is slightly
greater than the threshold, then the probability that every node in
the network is able to reconstruct the block quickly approaches
one. \\
\indent All of the aforementioned works
\cite{Eryil,Starobinski,Ghaderi} study the gains of network
coding as the system size grows while the coding window size is held
constant. In particular, they show that the decoding delay of RLNC
scales as $O(\ln(n))$ for a fixed coding window size as
$n\rightarrow \infty$.  {\it However it can be seen that when the
coding window size is held constant, the throughput of the system
goes to zero as the system becomes large because each user receives
a block of $k$ packets in $O(\ln(n))$ time slots}. Therefore, it is
important to study the system when $k$ is scaled as a function of
$n$. In this paper, we investigate the decoding delay when the coding window size scales as a function of the network size. We observe that there is a phase transition with respect to broadcast throughput such that when the coding window size $k$ scales asymptotically slower than $\ln(n)$, the broadcast throughput converges to zero. On the other hand, scaling $k$ asymptotically faster than $\ln(n)$ leads to the broadcast throughput approaching $(1-p)$. This phase transition is fundamentally different from that observed in \cite{Starobinski} which relates to the decoding delay when $k$ is fixed while the phase transition observed in our work relates to the nature of scaling of $k$ as a function of network size $n$.\\
The following example motivates
our investigation of the throughput-decoding
delay tradeoff of RLNC:\\
\begin{example}
Consider a single source broadcasting blocks of $k$ packets to $n$ users in a rateless transmission. Each packet is a vector of length $m$ over a finite field $\mathbb{F}_d$. The channel between the source and each of the users is a time invariant erasure channel with probability of erasure $p$. In each time slot, the source broadcasts a random linear combination of $k$ packets. Using random linear coding arguments introduced by Ho et al.~\cite{Ho}, for a large enough field size $d$, it is sufficient for the users to receive approximately $k$ coded packets to be able to decode the block. \\
\indent
Let $r[t]$ represent the probability that any given
user receives at least $k$ packets in $t\ge k$ time
slots. Then,
$r[t]=\sum_{l=k}^{t}{t \choose l}(1-p)^l
p^{t-l}$. Here $r[t]$ represents the fraction
of users that have successfully decoded $k$ packets by time
$t$.\\
To compare the behavior of $r[t]$ as a function of $t$ for different
values of $k$, we define a normalized time variable,
$s=\frac{t-k}{k}$. Accordingly, we define $r'[s]=r[ks+k]$, which can
be interpreted as the fraction of users that have successfully
decoded a single packet in a block of $k$ packets by $s$ time slots.
The comparison of $r'[s]$ for different $k$ allows us to see, in a
normalized time scale, the fraction of users that can decode an
equivalent of a single packet from a batch of $k$.
\begin{figure}[ht]
\centering
\includegraphics[width=3in]{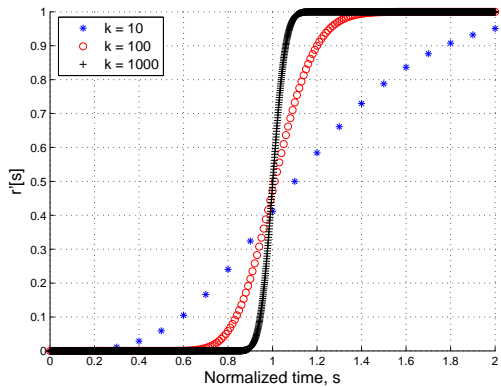}
\caption{Fraction of users that have successfully decoded a
single packet in a block of $k$ packets in $s$ time slots, $r'[s]$
as a function of $s$ for $p =1/2$} \label{fig:example}
\end{figure}
We numerically evaluate $r'[s]$ as a function of $s$ for different
values of $k$ as shown in the Figure \ref{fig:example} for the case
where $p = 0.5$. It can be seen from the graph that for $k=10$, a
large fraction of users are served within a short duration and then
the source takes a relatively longer time to serve the remaining small
fraction of users towards the end of the transmission of the current
block of $k$ packets. On increasing $k$ to $k=100$, the graph becomes
sharper indicating that the source serves a larger fraction of users
in a shorter duration and takes lesser time to serve a smaller
fraction of users towards the end of the transmission.

Ideally, we would like all the users to complete decoding together
for an increase in throughput. This can be achieved by increasing
$k$ indefinitely as observed from Figure \ref{fig:example}. However,
this causes the decoding delay to increase indefinitely as well.
Hence, it is important to understand the throughput-delay tradeoff
as $k$ scales as a function of $n$.
\end{example}

\section{System Model}
\label{sec:model}

\begin{figure}[ht]
\centering
\includegraphics[width=3in]{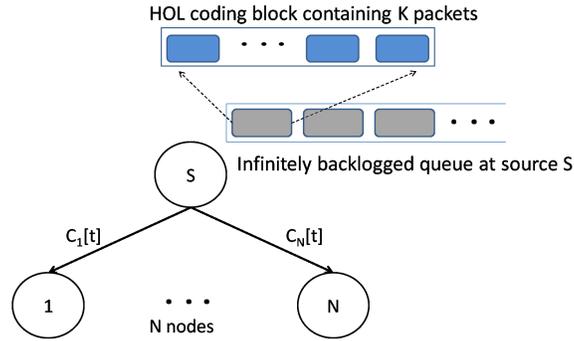}
\caption{A single source broadcasting to $n$ users over erasure
channels with probability of erasure, $p$ in each time slot.}
\label{fig:model}
\end{figure}

\begin{figure}[ht]
\centering
\includegraphics[width=2.3in]{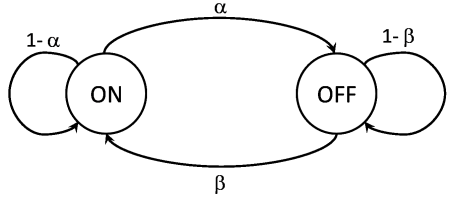}
\caption{A time correlated erasure channel model. When the channel is in state \textsc{on} the transmission is successful with probability $1$ while in state \textsc{off} the transmission fails with probability $1.$}
\label{fig:channel}
\end{figure}

In this work, we study the basic wireless broadcast scenario
depicted in Figure \ref{fig:model} that models the characteristics
of cellular or satellite systems and serves as the fundamental
building block for more general networks.
In particular, we consider a single source node, $S,$ broadcasting
an infinite backlog of common information to $n$ users over independent
time-varying erasure channels. The data is encapsulated into
\emph{packets}, each represented as a vector of length $m$ over a
finite field $\mathbb{F}_d$. We assume a time-slotted operation of
the system with $C_i[t] \in \{0, 1\}$ denoting the state of User
$i'$s channel in slot $t$. A single packet may be broadcast in each time slot by
the source and the transmission to the $i^{th}$ user is successful
only if $C_i[t]=1$. Let $\textbf{C}[t]=\left[C_1[t],\ldots,C_n[t]\right]$ be the $n-$dimensional vector of channel states of all users in slot $t$. We refer to $\textbf{C}[t]$ as the channel state of our system. $\textbf{C}[t]\in \mathcal{S}^n$ where $\mathcal{S}=\{0,1\}.$ For simplicity, we assume that all channels are independent and identically distributed. We assume a time correlated erasure channel model between the source and each of the users that is defined next.
\begin{definition}[Time Correlated Erasure Channel Model]
\label{def:channel} In each time-slot, user $i'$s channel is in one of two states as shown in the Figure \ref{fig:channel}. When the channel is in the \textsc{on} state, $C_i[t]=1$ with probability $1$ while in \textsc{off} state, $C_i[t]=0$ with probability $1.$ The state of the channel evolves as a Markov chain with transition probabilities $\alpha$ and $\beta.$ Let $p$ denote the steady state probability of an erasure over the channel. Then $p=\frac{\alpha}{\alpha+\beta}$. By setting $\alpha+\beta=1$, we can obtain, as a special case, the time invariant erasure channel model, where $C_i[t]=1$ with probability $1-p$ and $C_i[t]=0$ with probability $p$.
\end{definition}

We consider the class of block coding strategies employed by the source, where data is transferred in blocks of $k$ packets.
Specifically, the source can start transmitting the next block only
if the previous block is successfully transferred to all $n$
receivers. Moreover, we focus on the Random Linear Network Coding
(RLNC) strategy that is defined next.
\begin{definition}[Random Linear Network Coding (RLNC)]
\label{def:rlnc} In each time slot, the source transmits a random
linear combination of the $k$ packets in the Head-of-line (HOL)
coding block (see Figure~\ref{fig:model}) with coefficients of combination chosen uniformly at random from the field $\mathbb{F}_d$. In what follows, we refer
to $k$ as the {\it coding window (or block) size} of RLNC.
\end{definition}
Using random linear coding arguments~\cite{Ho}, the probability that the original $k$ packets can be decoded from any $k$ encoded packets formed using the RLNC scheme is equal to $\displaystyle \left(1-\frac{1}{d^k}\right)\left(1-\frac{1}{d^{k-1}}\right)\ldots\left(1-\frac{1}{d}\right)$ which is bounded below by $\left(1-\frac{1}{d-1}\right).$ Therefore, for a large enough field size $d$, it is sufficient for the users to receive approximately $k$ coded packets to be
able to decode the block. Hence, we assume that, under the RLNC scheme, the source continues to transmit encoded packets of the current block until each user successfully receives $k$ linear combinations.\\
It has been shown in \cite{Lun} that
random linear network coding is capacity achieving for multicast
connections in an unreliable network setting. That is, for $k$ sufficiently large, under the coding scheme
defined in Definition \ref{def:rlnc}, the (broadcast) capacity of
our system is $(1-p)$. Next, we define the two metrics of interest in our analysis, namely
throughput and decoding-delay.
\begin{definition}[(Broadcast) Throughput]
We let $R[t]$ denote the number of packets transmitted by the source in a total
of $t$ slots. Then, the (broadcast) throughput for a given
$n$ and $k$ under RLNC scheme, denoted as $\eta(n,k)$, is the
long-term average number of successfully transferred data packets to
all $n$ users. Hence, we have
\begin{equation}
\eta(n,k) = \lim_{t \rightarrow \infty} \frac{R[t]}{t}
\end{equation}
\end{definition}

\begin{definition}[Decoding-Delay]
\label{def:delay} We let $Y^{(j)}_i$ denote the number of time slots it
takes for the $i^{th}$ user to decode the $j^{th}$ block of $k$ packets
under the RLNC scheme. The decoding delay of the $j^{th}$ block, for a given $n$ and $k,$ under
the RLNC scheme, denoted as $U^{(j)}$, is the time required to
transmit all packets of the head-of-line (HOL) block to all the
users. Hence, we have
\begin{equation}
\label{eq:delaycorr}
U^{(j)} = \max_{1\le i \le n} Y^{(j)}_i
\end{equation}
\end{definition}
Recall that the source transmits linear combinations of the current block until each user successfully receives $k$ combinations. Then $Y^{(j)}_i$ is the time it takes for $k$ successful transmissions on User $i'$s channel and can be written as the sum of durations between each of the $k$ successful transmissions. Hence,
\begin{equation}
\label{eq:sumCLT}
Y^{(j)}_i=\sum_{h=1}^k X^{(j)}_{ih},
\end{equation}
where $X^{(j)}_{ih}$ is the duration between the ${(h-1)}^{th}$ and $h^{th}$ successful transmission on User $i'$s channel for the $j^{th}$ coding block.

To understand the importance of scaling $k$ as a function of $n$ to guarantee a non-vanishing throughput, consider a simple time invariant channel model, i.e, let $C_i[t]$ be a Bernoulli random variable with $p$ being the probability that $C_i[t] = 0$  in any given time slot $t$. Owing to the block transmission structure together with the independence of the channel states across time, the decoding delay of the block transmissions $\{U^{(j)}\}_{j\ge 1}$ are independent and identically distributed. This allows us to model the RLNC operation as
a renewal process with renewals at the start of each coding block
formation and $\{U^{(j)}\}_{j\ge 1}$ being the sequence of inter-renewal intervals. Let the random variable $U$ be identically distributed as $\{U^{(j)}\}_{j\ge 1},$ and, $\mathbb{E}[U^{(j)}]=\mathbb{E}[U]$ for all $j.$ Now, by defining a constant reward of $k$ acquired in
each renewal interval, we can utilize the main result from renewal
theory \cite{Ross} to write:
\begin{equation} \label{eq:gput}
\eta(n,k)= \lim_{t \rightarrow \infty} \frac{R[t]}{t}  = \frac{k}{\mathbb{E}[U]} \\
\end{equation}
\indent Under the time invariant erasure channel model, it is easy
to see that for each $i$ and $j$, $Y^{(j)}_i$ is a negative binomial random variable of order $k$ and success probability $(1-p)$. The exact expression for $\mathbb{E}[U]$ is as follows,
\begin{equation}
\label{eq:delay}
\mathbb{E}[U] = k+\sum_{t=k}^{t=\infty}\Bigg[1-\Bigg(\sum_{\tau=k}^{\tau=t} {\tau - 1 \choose k-1} p^{(\tau-k)} q^{k}\Bigg)^n\Bigg],
\end{equation}
where ${s\choose m}$ gives the number of size $m$ combinations of $s$ elements and $q \triangleq (1-p)$. \\
\indent It can be seen from the above expression that, when $k$ is a
constant independent of $n$, the mean decoding delay $\mathbb{E}[U]$
increases with $n$. Thus, for any fixed $k,$ $\eta(n,k)$ in
(\ref{eq:gput}) goes to
zero as $n$ approaches $\infty$. However, the exact expression for  $\mathbb{E}[U]$ is difficult to simplify further.\\
\indent In the next section, we provide throughput and delay analysis of RLNC for the time-correlated channel model. By identifying a suitable renewal process with reward $R(t),$ we express throughput in terms of $\mathbb{E}[U^{(j)}]$ and $k$ using the main result in renewal theory for renewal-reward processes~\cite{Ross}. However, explicit characterization of $\mathbb{E}[U^{(j)}]$ is difficult. Instead, we derive upper and lower bounds on $\mathbb{E}[U^{(j)}]$ which enable us to understand the scaling of $k$ as a function of $n$ to guarantee a non-vanishing throughput. Further, we compute the mean and variance of $U^{(j)}$ when $k$ scales faster than $\ln(n)$.

\section{Analysis of RLNC for Time Correlated Erasure Channels}
\label{sec:corranalysis}
\indent Under the time correlated channel model (when $\alpha+\beta\ne1$ in Definition~\ref{def:channel}), the RLNC operation is not a renewal process after each coding block transmission and (\ref{eq:gput}) is not valid. Hence, in order to express the broadcast throughput in terms of the decoding delay, we model the RLNC operation over time correlated channels as a semi-Markov process and identify a suitable renewal-reward process. By deriving upper and lower bounds on the decoding delay and hence, the throughput, we show that a phase transition in the throughput performance of our system occurs at the block-length scaling rate of $k=\Theta(\ln(n))$. This result is summarized in the following theorem:
\begin{theorem}
\label{thm:main}
Under the RLNC scheme, the broadcast throughput $\eta(n,k)$ of our system, when $\alpha,\beta >0,$ and $\alpha+\beta\ne 2,$ can be characterized as follows,

\begin{enumerate}
\item If $k=o(\ln(n)),$ then
\begin{equation}
\lim_{n\rightarrow \infty} \eta(n,k)= 0.
\end{equation}
\item If $k=\omega(\ln(n)),$ then
\begin{equation}
\lim_{n\rightarrow \infty} \eta(n,k)=1-p.
\end{equation}
\item Furthermore, if $k=\Theta(\ln(n)),$ then
\begin{equation}
\liminf_{n\rightarrow \infty} \eta(n,k)\ge r(1-p),
\end{equation}
where $0<r<1$ is given in (\ref{eq:boundr2}).
\end{enumerate}
\end{theorem}
\indent In what follows, we develop the mathematical model required to prove the above result. We also characterize the asymptotic distribution of the decoding delay for the scaling regime of $k=\omega(\ln(n))$ and derive approximate expressions for the mean and variance of the decoding delay using extreme value theory. These expressions reveal the effect of channel correlation on the throughput and delay performance of RLNC.
\subsection{Throughput and Delay Analysis}
\label{sec:corranalysisA}
Let the random variable $\textbf{E}^{(j)}\in \mathcal{S}^n$ denote the channel state of the system at the end of the $j^{th}$ block transmission. Let $T^{(j)}\in \mathbb{Z}^+$ denote the time slot of the completion of the $j^{th}$ block transmission. Set $T^{(0)}=0$. Then, the stochastic process $\{\textbf{E}[t],t\ge 0\},$ where, for each $j\ge 0$, $\textbf{E}[t]=\textbf{E}^{(j)}$ for $t$ in the interval $T^{(j)}\le t < T^{(j+1)},$ is a semi-Markov process where the state transitions occur at the end of each block transmission. Let $\textbf{E}^{(0)}=\textbf{1}$, i.e, the process starts with all the user channels being \textsc{on}. We have that $U^{(j)} = T^{(j)}-T^{(j-1)},j\ge 1$ is the time between successive state transitions. Note that $U^{(j)}$ is the decoding delay of the $j^{th}$ block transmission and is a non-negative random variable that depends only on the current state and the next state of the Markov chain. The embedded Markov chain $\{\textbf{E}^{(j)}, j\ge 0\}$ is an irreducible finite state ($2^n$ states) Markov chain when $\alpha\ne 0$ and $\beta\ne 0.$ Also, the embedded Markov chain is aperiodic when $\alpha\ne 1$ or $\beta\ne 1.$\footnote{The cases when $1)$ $\alpha=\beta=1$ and $2)$ $\alpha=0$ or $\beta=0$ are trivial since the channel becomes deterministic within a finite expected number of time-slots.} Let the random variable $\textbf{B}^{(j)}\in \mathcal{S}^n$ denote the channel state of the system at the beginning of $j^{th}$ block transmission.  \\
\indent The time epochs of return to the channel state $\textbf{s}=\textbf{1}$ in the semi-Markov process form a renewal process.  The source completes transmission of a random number of blocks in each inter-renewal interval. Let the random variable $W$ denote the length of one such inter-renewal interval. Let $M$ denote the random number of block transmissions completed in this interval. Therefore, we have the identity $\displaystyle W=\sum_{j=1}^M U^{(j)}.$ \\
Now, $M$ represents the number of transitions in the embedded Markov chain between consecutive visits to state $\textbf{1}$. Since the embedded Markov chain is an irreducible aperiodic finite state Markov chain, we have that the steady state probabilities of each state $\textbf{s}$ in the embedded chain, $\pi_{\textbf{s}}$ is strictly positive. Hence $\mathbb{E}[M] = \frac{1}{\pi_{\textbf{s}}} < \infty$ \cite{Ross}.  \\
\indent Now, since $\mathbb{E}[M]<\infty$, by defining a reward of $1$ for each packet transmitted in a renewal interval, we can utilize the main result from renewal
theory \cite{Ross} to write:
\begin{equation}
\label{eq:corrgput}
\eta(n,k)= \lim_{t \rightarrow \infty} \frac{R(t)}{t}  = \frac{k\mathbb{E}[M]}{\mathbb{E}[W]}. 
\end{equation}
The random variable $W$ depends on the channel state of the system at the beginning of each of the $M$ block transmissions. Also, owing to the time correlated channel process, the random variables $\{U^{(j)}\}_{j=1}^M$ are not ${\it i.i.d}$. An explicit characterization of $\mathbb{E}[W]$ in terms of $\{U^{(j)}\}$ is, therefore, difficult. Instead, we obtain upper and lower bounds on $\mathbb{E}[W].$ Recall that $U^{(j)},$ $j\ge 1$ is the decoding delay of the $j^{th}$ block. Now using (\ref{eq:delaycorr}) and (\ref{eq:sumCLT}), we can bound $U^{(j)}$ as follows,
\begin{align}
\label{eq:holdtime}
\max_{1\le i\le n} X^{(j)}_{i2} &\le U^{(j)} \le \max_{1\le i\le n}  X^{(j)}_{i1} +  \max_{1\le i\le n} \left(X^{(j)}_{i2}+\ldots+X^{(j)}_{ik}\right) \quad a.s.
\end{align}
Note that, for all $h\ge 2,$ $X^{(j)}_{ih}$ is the number of time-slots between the ${(h-1)}^{th}$ and $h^{th}$ successful transmission given that the channel state of User $i$ in the time slot of reception of ${(h-1)}^{th}$ linear combination is \textsc{on}. Therefore, the random variables $\{X^{(j)}_{ih}\}$ $i=1,\ldots,n,$ $\forall j\ge 1,$ $\forall h\ge 2$ are ${\it i.i.d}$. We also have that, for all $j\ge 1$ and $h\ge 2,$ $X^{(j)}_{ih}$ is independent of $\textbf{E}^{(1)},\ldots,\textbf{E}^{(j-1)}.$ Also, note that $\displaystyle X^{(j)}_{i1},$ $j\ge 1$ is not independent of $\textbf{E}^{(j-1)}.$\\
\indent Now, the identity $\displaystyle W=\sum_{j=1}^M U^{(j)}$ along with (\ref{eq:holdtime}) enables us to bound  $\mathbb{E}[W]$ as follows,
\begin{align}
\label{eq:corrfullbds1}
 \mathbb{E}[W] &\ge \mathbb{E} \left[ \sum_{j=1}^M \max_{1\le i\le n} X^{(j)}_{i2} \right] \\
\label{eq:corrfullbds2}
\mathbb{E}[W] &\le \mathbb{E}\left[\sum_{j=1}^M \max_{1\le i\le n} X^{(j)}_{i1}\right] + \mathbb{E}\left[ \sum_{j=1}^M \max_{1\le i\le n} \left( X^{(j)}_{i2}+\ldots+X^{(j)}_{ik}\right)\quad \right]
\end{align}
To prove Theorem~\ref{thm:main}, we further bound the lower and upper bounds on $\mathbb{E}[W]$ in (\ref{eq:corrfullbds1}) and (\ref{eq:corrfullbds2}), respectively. First, we state a useful lemma~\cite{Shrader}:
\begin{lemma}(From \cite{Shrader}, pg. 7)
\label{lem:main}
Let $\chi_i, i =1,2,\ldots,n$ be identically distributed random variables. Then for any $\tau>0$, we have that
\begin{equation}
\label{eqn:ineq1}
\mathbb{E}[\max_{1\le i \le n} \chi_i] \le \frac{1}{\tau}\Bigg(\ln(n)+ \ln(\mathbb{E}[e^{\tau\chi_1}]))\Bigg)
\end{equation}
\end{lemma}
Note that the above bound is useful only when $\tau$ lies within the radius of convergence of the moment generating function (m.g.f) of $\chi_1$.\\
Next, we prove a few lemmas that will be useful in the proof of Theorem~\ref{thm:main}. Recall that $\textbf{B}^{(j)}\in \mathcal{S}^n$ denotes the channel state of the system at beginning of $j^{th}$ block transmission. For any fixed $n,$ $|\mathcal{S}^n|$ is finite. Hence, $\displaystyle \max_{\textbf{s}\in \mathcal{S}^n} \mathbb{E}\left[\max_{1\le i\le n}X^{(j)}_{ih}|\textbf{B}^{(j)}=\textbf{s}\right]$ exists and can be bounded as follows,
\begin{lemma}
\label{lem:singleub}
 There exist constants $\mu_0$ and $\tau_0$ such that, for all $j$ and $h,$
\begin{align*}
\max_{\textbf{s}\in \mathcal{S}^n} \mathbb{E}\left[\max_{1\le i\le n}X^{(j)}_{ih}|\textbf{B}^{(j)}=\textbf{s}\right] \le \frac{1}{\tau_0}\left(\ln(n)+\mu_0\right).
\end{align*}
Also, $\displaystyle \mathbb{E}\left[\max_{1\le i\le n}X^{(j)}_{ih}\right] < \infty$ for any $j,$ $h\in\{1,\ldots,k\}.$ \\
\end{lemma}
\begin{IEEEproof}
Fix a $\tau_0< \ln(\frac{1}{1-\beta}).$ Then $\mathbb{E}\left[e^{\tau_0X^{(j)}_{ih}}\right]$ exists and is finite. Let $B_i^{(j)}$ be the channel state of the $i^{th}$ user at the beginning of the $j^{th}$ block transmission. Let $\displaystyle \mu_0=\max_{c\in\{0,1\}}\mathbb{E}\left[e^{\tau_0X^{(j)}_{ih}}|B_i^{(j)}=c\right].$ Now, since the channel state of any user is independent of other users' channels, $\displaystyle \mathbb{E}\left[e^{\tau_0X^{(j)}_{ih}}|\textbf{B}^{(j)}=\textbf{s}\right]\le \mu_0$ for all $i,$ $j,$ and, $h\in\{1,\ldots,k\}.$
\begin{align*}
\mathbb{E}\left[\max_{1\le i\le n} X^{(j)}_{ih}|\textbf{B}^{(j)}=\textbf{s}\right] &\le  \mathbb{E}\left[\frac{1}{\tau_0}\ln\left(\max_{1\le i\le n} e^{\tau_0X^{(j)}_{ih}}\right)|\textbf{B}^{(j)}=\textbf{s}\right]\\
&\le   \frac{1}{t}\ln\left(\mathbb{E}\left[e^{\tau_0X^{(j)}_{1h}}+\ldots+e^{\tau_0X^{(j)}_{nh}}|\textbf{B}^{(j)}=\textbf{s}\right]\right) \\
\label{eq:corrubpart1}
&\le   \frac{1}{\tau_0}\left(\ln(n)+\ln(\mu_0)\right)
\end{align*}
\end{IEEEproof}
In the next lemma, we find an upper bound on the expected value of $\displaystyle \sum_{j=1}^M \max_{1\le i\le n}X^{(j)}_{i1}$:
\begin{lemma}
Let $\mu=\frac{1}{\tau_0}\left(\ln(n)+\ln(\mu_0)\right)$ where $\tau_0$ and $\mu_0$ are defined in the previous lemma. Then,
\label{lem:corrub1}
\begin{equation}
\mathbb{E}\left[\sum_{j=1}^M \max_{1\le i\le n}X^{(j)}_{i1}\right] \le \mathbb{E}[M]\mu.
\end{equation}
\end{lemma}
\begin{IEEEproof}
Let  $\displaystyle \xi^{(j)}=\max_{1\le i\le n}X^{(j)}_{i1}$. Recall that $\textbf{E}^{(j)} \in \mathcal{S}^n$ is the channel state of the system at the end of the $j^{th}$ block transmission. Let $\mathcal{F}_j$ be the smallest $\sigma-$field of events containing the  $\sigma-$fields generated by the random variables $\{X^{(1)}_{ih},\ldots,X^{(j)}_{ih}, i=1,\ldots,n, h = 1,\ldots, k\}$ and $\textbf{E}^{(1)},\ldots,\textbf{E}^{(j)}$. Let $Z_0=0.$ Define $\displaystyle Z^{(m)}=\sum_{j=1}^m(\xi^{(j)}-\mu^{(j)})$ where $\mu^{(j)}=\mathbb{E}[\xi^{(j)}|\mathcal{F}_{j-1}].$ Also,
\begin{align*}
\mu^{(j)}&=\mathbb{E}[\xi^{(j)}|\mathcal{F}_{j-1}]\\
&=\sum_{\textbf{s}\in \mathcal{S}} \mathbb{E}[\xi^{(j)}|\textbf{B}^{(j)}=\textbf{s},\mathcal{F}_{j-1}]P(\textbf{B}^{(j)}=\textbf{s}|\mathcal{F}_{j-1})\\
&=\sum_{\textbf{s}\in \mathcal{S}} \mathbb{E}[\xi^{(j)}|\textbf{B}^{(j)}=\textbf{s}]P(\textbf{B}^{(j)}=\textbf{s}|\mathcal{F}_{j-1})\\
&\le \mu,
\end{align*}
where the last inequality follows from Lemma~\ref{lem:singleub}. Then $\mathbb{E}[\xi^{(j)}]=\mathbb{E}[\mu^{(j)}] \le \mu.$ We have that $\{Z^{(m)}\}$ is a martingale by definition. Also, $M$ is a stopping time  with respect to the filtration $\{\mathcal{F}_{j}\}$ with $\mathbb{E}[M]<\infty$ and
\begin{align*}
\mathbb{E}[|Z^{(m+1)}-Z^{(m)}||\mathcal{F}_{m}] &= \mathbb{E}[|\xi^{(m+1)}-\mu^{(m+1)}||\mathcal{F}_{m}]\\
& \le \mathbb{E}[\xi^{(m+1)}|\mathcal{F}_{m}]+\mu^{(m+1)} \\ &= 2\mu^{(m+1)} \le 2\mu < \infty.
\end{align*}
Hence, using the optional stopping theorem (Theorem 7.2.2 in \cite{Ross}), we have $\mathbb{E}[Z^{(M)}]=\mathbb{E}[Z^{(1)}]=0.$ Therefore,
\begin{align*}
0=\mathbb{E}[Z^{(M)}]&=\mathbb{E}\left[\sum_{j=1}^M\xi^{(j)}\right]-\mathbb{E}\left[\sum_{j=1}^M \mu^{(j)}\right] \\
&\ge \mathbb{E}\left[\sum_{j=1}^M\xi^{(j)}\right]-\mathbb{E}\left[M \mu\right].
\end{align*}
Thus, we have $\displaystyle \mathbb{E}\left[\sum_{j=1}^M\xi^{(j)}\right]\le \mathbb{E}[M]\mu$ as required.
\end{IEEEproof}
In the next lemma, we find a lower bound on the expected value of $\displaystyle \max_{1\le i\le n} X^{(j)}_{i2}:$
\begin{lemma}
\label{lem:corrlb}
Let $\lambda=-\ln(1-\beta).$ Then, for all $j,$ we have
\begin{equation}
 \mathbb{E}\left[\max_{1\le i\le n} X^{(j)}_{i2}\right]\ge 1 +  \frac{\alpha}{\lambda}\left[\ln(n)+\gamma\right].
 \end{equation}
\end{lemma}
\begin{IEEEproof}
We have that $P(X^{(j)}_{i2}=1)=1-\alpha$ and $P(X^{(j)}_{i2} = u)=\alpha\beta(1-\beta)^{u-2},$ for all $u\ge 2.$ Hence,
\begin{align*}
P(X^{(j)}_{i2}\le u)&=(1-\alpha)+\sum_{v=2}^u\alpha\beta(1-\beta)^{v-2}\\
&=1-\alpha(1-\beta)^{u-1}
\end{align*}
Therefore, $\displaystyle P(\max_{1\le i\le n} X^{(j)}_{i2}\le u) = (1-\alpha(1-\beta)^{u-1})^n \quad \forall u \ge 1$ and $\displaystyle \mathbb{E}\left[\max_{1\le i\le n} X^{(j)}_{i2}\right]$ is given as
\begin{align*}
\mathbb{E}\left[\max_{1\le i\le n} X^{(j)}_{i2}\right]&=1+\sum_{u=1}^\infty \left(1- \left[1-\alpha e^{-\lambda(u-1)}\right]^n\right)\\
&\ge 1 + \int_0^\infty \left(1-(1-\alpha e^{-\lambda x})^n\right) dx
\end{align*}
By setting $w=1-\alpha e^{-\lambda x} \in [0,1)$ in the above integral, we have the following inequalities,
\begin{align*}
\mathbb{E}\left[\max_{1\le i\le n} X^{(j)}_{i2}\right]&\ge  1 + \int_{1-\alpha}^1 \frac{1-w^n}{\lambda(1-w)} dw \\
&= 1 + \frac{1}{\lambda}\int_{1-\alpha}^1 \sum_{z=0}^{n-1} w^z dw\\
&=  1 + \frac{1}{\lambda} \sum_{z=0}^{n-1} \int_{1-\alpha}^1 w^z dw\\
&= 1 + \frac{1}{\lambda} \sum_{z=0}^{n-1} \frac{1}{z+1} - \frac{1}{\lambda}\sum_{z=0}^{n-1} \frac{(1-\alpha)^{z+1}}{z+1}\\
&\ge  1 + \frac{1}{\lambda} \sum_{z=0}^{n-1} \frac{1}{z+1} - \frac{1}{\lambda}\sum_{z=0}^{n-1} \frac{1-\alpha}{z+1}\\
&\ge  1 +  \frac{\alpha}{\lambda}\left[\ln(n)+\gamma\right].
\end{align*}
\end{IEEEproof}
Next, we prove Theorem~\ref{thm:main} using the above lemmas and Wald's first equation to bound the lower and upper bounds in (\ref{eq:corrfullbds1}) and (\ref{eq:corrfullbds2}), respectively.
\begin{IEEEproof}(Proof of the Theorem~\ref{thm:main})\\
{\it Proof of $(a):$} First, we prove that $k=o(\ln(n))$ leads to a vanishing throughput. As mentioned in Lemma~\ref{lem:corrub1}, $M$ is a stopping time with respect to the filtration $\{\mathcal{F}_j\}$. We have that $\displaystyle \{\max_{1\le i\le n} X^{(j)}_{i2}\}_{j\ge 1}$ are identically distributed and have a finite mean. Also, $\displaystyle \max_{1\le i\le n} X^{(j)}_{i2}$ is independent of $\mathcal{F}_{j-1}$. Thus, using Wald's first equation and Lemma~\ref{lem:corrlb}, we can lower bound $\mathbb{E}[W]$ (cf. (\ref{eq:corrfullbds1})) as follows,
\begin{align*}
\mathbb{E}[W] &\ge \mathbb{E}[M]\mathbb{E}\left[\max_{1\le i\le n} X^{(1)}_{i2}\right]\\
&\ge \mathbb{E}[M]\left(1 +  \frac{\alpha}{\lambda}\left[\ln(n)+\gamma\right]\right).
\end{align*}
Consequently, using (\ref{eq:corrgput}), we can upper-bound the throughput  as follows,
\begin{align}
\eta(n,k) &\le \frac{k\mathbb{E}[M]\lambda}{\mathbb{E}[M]\left(\lambda +  \alpha\left[\ln(n)+\gamma\right]\right)}\\
&= \frac{k\lambda}{\lambda + \alpha\left[\ln(n)+\gamma\right]}.
\end{align}
Clearly, when $k=o(\ln(n))$, the upper bound, and hence, $\eta(n,k)$ decreases to zero, as $n\rightarrow \infty.$ \\
{\it Proof of $(b):$} Next, we consider the case when $k=\omega(\ln(n))$. We let $k=f(n)\ln(n)+1$ for some function $f(n)>0$ such that $\displaystyle \lim_{n\rightarrow \infty}f(n)=\infty.$ Let $\hat{Y}^{(j)}_i =  X^{(j)}_{i2}+\ldots+X^{(j)}_{ik}$. Let $\hat{\mu}$ and $\hat{\sigma}^2$ be the mean and variance of $\hat{Y}^{(j)}_i$. Then,
\begin{align*}
\hat{\mu}&=(k-1)(1+\frac{\alpha}{\beta})=(k-1)\frac{1}{1-p},\\
\hat{\sigma}^2&=\frac{(k-1)\alpha[2-(\alpha+\beta)]}{\beta^2}.
\end{align*}
As mentioned in Lemma~\ref{lem:corrub1}, $M$ is a stopping time with respect to the filtration $\{\mathcal{F}_j\}$. We have that $\displaystyle \{\max_{1\le i\le n}\hat{Y}^{(j)}_i\}_{j\ge 1}$ are identically distributed and have a finite mean. Also, $\displaystyle \max_{1\le i\le n}\hat{Y}^{(j)}_i$ is independent of $\mathcal{F}_{j-1}.$ Using Wald's first equation and Lemma~\ref{lem:corrub1}, we can rewrite (\ref{eq:corrfullbds2}) as follows,
\begin{align}
\label{eq:corrub2}
\mathbb{E}[W] &\le \mathbb{E}\left[\sum_{j=1}^M \max_{1\le i\le n} X^{(j)}_{i1}\right] + \mathbb{E}[M]\hat{\mu} +  \hat{\sigma} \mathbb{E}\left[\sum_{j=1}^M \max_{1\le i\le n} \frac{\hat{Y}^{(j)}_i-\hat{\mu}}{\hat{\sigma}}\right] \nonumber \\
&\le \mathbb{E}[M]\left(\mu + \hat{\mu} +  \hat{\sigma}\mathbb{E}\left[\max_{1\le i\le n} \frac{\hat{Y}^{(j)}_i-\hat{\mu}}{\hat{\sigma}}\right]\right).
\end{align}
Next, we upper bound $\displaystyle \mathbb{E}\left[\max_{1\le i\le n} \frac{\hat{Y}^{(j)}_i-\hat{\mu}}{\hat{\sigma}}\right]$ by appealing to Lemma~\ref{lem:main}. Choose $\tau=b\sqrt{\ln(n)}$ where $b$ is a constant such that $0 < \tau\le\hat{\sigma}\ln(\frac{1}{1-\beta})$. For such a $\tau$,
\begin{align*}
\mathbb{E}\left[\exp\left(\tau\frac{\hat{Y}^{(j)}_i-\hat{\mu}}{\hat{\sigma}}\right)\right]&= \exp\left(\frac{-\tau\hat{\mu}}{\hat{\sigma}}\right)\left(\frac{\exp\left(\frac{\tau}{\hat{\sigma}}\right)\left[1-\alpha\left(1-\exp\left(\frac{\tau}{\hat{\sigma}}\right)\right)-\exp\left(\frac{\tau}{\hat{\sigma}}\right)\left(1-\beta\right)\right]}{1-\exp\left(\frac{\tau}{\hat{\sigma}}\right)\left(1-\beta\right)}\right)^{k-1}\\[5pt]
&= r(n)^{f(n)\ln(n)},
\end{align*}
where $r(n)$ is defined as follows
\begin{equation*}
r(n)=\left(\frac{\exp\left(\frac{d_1}{\sqrt{f(n)}}\right)\left[1-\alpha\left(1-\exp\left(\frac{d_2}{\sqrt{f(n)}}\right)\right)-\exp\left(\frac{d_2}{\sqrt{f(n)}}\right)\left(1-\beta\right)\right]}{1-\exp\left(\frac{d_2}{\sqrt{f(n)}}\right)\left(1-\beta\right)}\right),
\end{equation*}
 with $\displaystyle d_1 = \frac{-b\sqrt{\alpha}}{\sqrt{2-\alpha-\beta}}$ and $\displaystyle d_2=\frac{b\beta}{\sqrt{\alpha(2-\alpha-\beta)}}.$\\


Now, using Lemma~\ref{lem:main}, we obtain the following upper bound,
\begin{align}
\mathbb{E}\left[\max_{1\le i\le n} \frac{\hat{Y}^{(j)}_i-\hat{\mu}}{\hat{\sigma}}\right]&\le\frac{1}{\tau}\left(\ln(n)+\ln\left(\mathbb{E}\left[\exp\left(t\frac{\hat{Y}^{(j)}_i-\hat{\mu}}{\hat{\sigma}}\right)\right]\right)\right)\nonumber\\
\label{eq:corrub3}
&=\frac{1}{b}\sqrt{\ln(n)}\left(1+f(n)\ln\left(r(n)\right)\right)
\end{align}
Using (\ref{eq:corrub2}) and (\ref{eq:corrub3}), we can bound $\mathbb{E}[W]$ and $\eta(n,k)$ as follows,
\begin{align}
\mathbb{E}[W]&\le \mathbb{E}[M]\frac{\phi(n)}{1-p}, \nonumber \\
\label{eq:goodputcorr}
\eta(n,k)&\ge \frac{(1-p)(f(n)\ln(n)+1)}{\phi(n)},
\end{align}
where $\phi(n)$ is as follows,
\begin{equation*}
\phi(n)= \left(\frac{1-p}{\tau_0}\left(\ln(n)+\ln(\mu_0)\right)+f(n)\ln(n)+\frac{(1-p)\sqrt{f(n)}\sqrt{\alpha[2-(\alpha+\beta)]}}{b\beta}\ln(n)\left(1+f(n)\ln\left(r(n)\right)\right)\right).
\end{equation*}
 \\
Since, $f(n)\rightarrow\infty,$  as $n\rightarrow\infty,$ we have $r(n)\rightarrow 1$ as $n\rightarrow\infty.$ Let $d_3=\frac{\sqrt{\alpha[2-(\alpha+\beta)]}}{b\beta}.$ Then,
\begin{align*}
\lim_{n\rightarrow\infty}\frac{\phi(n)}{f(n)\ln(n)+1}-1&=d_3\lim_{n\rightarrow\infty}\sqrt{f(n)}\ln\left(r(n)\right)\\
&=d_3\lim_{n\rightarrow\infty}\sqrt{f(n)}(1-r(n))\ln\left(1-(1-r(n))\right)^\frac{1}{1-r(n)}\\
&=-d_3\lim_{n\rightarrow\infty}\sqrt{f(n)}(1-r(n)) \tag{i}\\
&=\frac{-d_3}{\beta}\lim_{n\rightarrow\infty}\sqrt{f(n)}\left(1-\left(1-\beta\right)\exp\left(\frac{d_2}{\sqrt{f(n)}}\right)-\right.\\
&\left.\exp\left(\frac{d_1}{\sqrt{f(n)}}\right)\left[
1-\alpha\left(1-\exp\left(\frac{d_2}{\sqrt{f(n)}}\right)\right)-\left(1-\beta\right)\exp\left(\frac{d_2}{\sqrt{f(n)}}\right)\right]\right)\\
&=0 \tag{ii},
\end{align*}
where (i) follows by noting that $\left(1-(1-r(n))\right)^\frac{1}{1-r(n)}\rightarrow e^{-1}$ as $n\rightarrow\infty$ and (ii) follows by $L'H\hat{o}spital's$ rule. Hence, from (\ref{eq:goodputcorr}) and the fact that $\eta(n,k)\le(1-p)$, we see that under the scaling regime of $k=\omega(\ln(n)),$ we have $\eta(n,k)\rightarrow (1-p)$ as $n\rightarrow\infty.$\\
{\it Proof of $(c):$} Finally, consider the case when $k=\Theta(\ln(n)).$ This is achieved by letting $f(n)=\hat{b}$ for some constant $\hat{b}$ independent of $n$ in the above analysis. Then, $r(n)$ is a constant independent of $n$ and
\begin{align}
\label{eq:boundr2}
\lim_{n\rightarrow\infty}\frac{\phi(n)}{f(n)\ln(n)+1}&=1+\frac{1-p}{\tau_0\hat{b}}+\frac{\sqrt{\alpha[2-(\alpha+\beta)]}}{b\beta\sqrt{\hat{b}}}\left(1+\hat{b}\ln\left(r(n)\right)\right)\\
&>1 \nonumber
\end{align}
and from (\ref{eq:goodputcorr}), we see that a constant fraction of the capacity is guaranteed, hence proving $(c)$.
\end{IEEEproof}

\subsection{Computing the Mean and Variance of Decoding Delay}
In Theorem~\ref{thm:main}, we showed that it is necessary to scale $k$ at least as $\ln(n)$ to guarantee a non-vanishing broadcast throughput. Next, we aim to obtain an accurate characterization of the decoding delay, $U^{(j)}$ in the scaling regime of $k=\omega(\ln(n))$. In what follows, we drop the superscript $(j)$ and let $U$ denote the decoding delay of our system under the RLNC scheme.\\ 
First consider the case of time-invariant channel model. We see that, for each $j,$ $Y^{(j)}_i$ in (\ref{eq:sumCLT}) is the sum of ${\it i.i.d}$ random variables. By appealing to the Central Limit Theorem, we know that, after suitable standardization, $Y^{(j)}_i$ converges in distribution to a standard normal random variable as $k\rightarrow \infty.$ It is also well-known that, the distribution of the maximum of $n$ normal random variables, after suitable standardization, converges weakly to the Gumbel distribution \cite{hadavid} as $n\rightarrow \infty$. Therefore, we expect the decoding delay, $U,$ to converge in distribution to the Gumbel distribution, after suitable standardization, as $n\rightarrow\infty$. This is shown to be true in \cite{anderson} for the general case of maxima of triangular arrays. We summarize the result for our setting in the following theorem,
\begin{proposition}(cf. \cite{anderson}, pg. 961)
\label{thm:EVT}
Let $\mu(k)=\frac{k}{1-p}$ and $\sigma^2(k)=k\frac{p}{(1-p)^2}.$ When $k=\omega(\ln(n)),$ we have
\begin{align}
\lim_{n\rightarrow\infty} P\left(\frac{U-\mu(k)}{\sigma(k)}\le a_nx+b_n\right) = \exp\left(-e^{-x}\right),
\end{align}
where $\displaystyle a_n  \sim \frac{1}{\sqrt{2\ln(n)}}$ and $b_n  \sim \sqrt{2\ln(n)}$ as $n\rightarrow\infty.$\\
\end{proposition}

In general, convergence in distribution does not imply convergence in moments. However, for the case of distributions belonging to the domain of attraction of the Gumbel distribution, convergence of moments holds true \cite{evt}. Let $\displaystyle \tilde{U}=\frac{U-\mu(k)}{\sigma(k)}.$ Then, $\forall r\ge 0,$ we have,
\begin{align}
\lim_{n\rightarrow\infty} \mathbb{E}\left[\frac{\tilde{U}-b_n}{a_n}\right]^r = \int_{-\infty}^{\infty} x^r d\exp\left(-e^{-x}\right).
\end{align}
Now, $\displaystyle \int_{-\infty}^{\infty} x d\exp\left(-e^{-x}\right) = \gamma$ where $\gamma$ is the Euler's constant and $ \displaystyle \int_{-\infty}^{\infty} x^2 d\exp\left(-e^{-x}\right) = \frac{\pi^2}{6}+\gamma^2.$ This enables to write the following approximate expressions for the mean and variance of decoding delay, when $n$ is large,
\begin{align}
\label{eq:mean}
\mathbb{E}[U] &\approx \frac{k}{1-p}+\frac{\sqrt{kp}}{1-p}\left(\sqrt{2\ln(n)}+\gamma\frac{1}{\sqrt{2\ln(n)}}\right),\\
\label{eq:var}
var[U]&\approx\frac{kp\pi^2}{12(1-p)^2\ln(n)}.
\end{align}
By using (\ref{eq:mean}) in (\ref{eq:gput}), we can verify that when $k=\omega(\ln(n)),$ the throughput converges to $1-p$ as $n\rightarrow\infty.$ In Section~\ref{sec:num}, using simulations (cf. Figure~\ref{fig:ctimelnN},~\ref{fig:ctimekn}), we show that the above approximate expressions are accurate even for small values of $n.$ \\
In \cite{Starobinski}, the authors show that, when $k$ is fixed,
\begin{align}
\lim_{n\rightarrow\infty} P\left(U\le \tilde{a}_n x + \tilde{b}_n\right) = \exp\left(-e^{-x}\right),
\end{align}
where $\tilde{a}_n = \frac{-1}{\ln(p)}$ and $\tilde{b}_n  \sim \ln(n)$ as $n\rightarrow\infty.$\\
Now, since $\tilde{a}_n$ is a constant, the asymptotic distribution of the decoding delay, $U,$ is concentrated around the point $\tilde{b}_n$ with a fixed variance for all $n$ and hence, there exists a phase transition with respect to the decoding delay, $U,$ such that there exists a threshold on the number of transmissions below which the probability that a block of coded packets can be recovered by all the nodes in the network is close to zero. On the other hand, if the number of transmissions is slightly greater than the threshold, then the probability that every node in
the network is able to reconstruct the block quickly approaches one. However, when $k=\omega(\ln(n)),$ $var[U]$ in (\ref{eq:var}) increases with $n$ and the phase transition phenomenon observed in \cite{Starobinski} becomes less evident.

For the case of correlated channels, we analyze the decoding delay for the case when the channel state of the system is $\textbf{s}=\textbf{1}$ in the time slot prior to the beginning of the new block transmission. Let the random variable $U$ denote this decoding delay as well. In this case too, the decoding delay for each user can be written as a sum of ${\it i.i.d}$ random variables. Hence, we have a similar proposition as Proposition~\ref{thm:EVT}. We can write the approximate expressions for mean decoding delay and variance of decoding delay, when $n$ is large, for both the channel models, concisely, as follows,
\begin{align}
\label{eq:approxcorrmean}
\mathbb{E}[U] &\approx \frac{k}{1-p}+\frac{\sqrt{kp}}{1-p}\sqrt{\left(\frac{2}{\alpha+\beta}-1\right)}(\sqrt{(2\ln(n))}+\frac{\gamma}{\sqrt{(2\ln(n))}})\\
var[U]&\approx\frac{kp\pi^2}{12(1-p)^2\ln(n)}\left(\frac{2}{\alpha+\beta}-1\right)
\end{align}
Note that the above expressions are the same as (\ref{eq:mean}) and (\ref{eq:var}) for the time invariant channel if we choose $\alpha+\beta=1$ and $p=\frac{\alpha}{\alpha+\beta}.$ Although (\ref{eq:approxcorrmean}) is derived under the assumption that the channel state of the system before the beginning of the block transmission is ${\bf s=1},$ we show, using simulations (cf. Figure~\ref{fig:ctimelnNcorr},~\ref{fig:ctimekncorr}) in Section~\ref{sec:num}, it is an accurate upper bound on the decoding delay performance of the actual system.

We can observe an interesting fact from expression (\ref{eq:approxcorrmean}): we see that the mean decoding delay decreases as the channel becomes more positively correlated, i.e, $1<\alpha+\beta\rightarrow 2$ and increases as the channel becomes more negatively correlated,  i.e, $\alpha+\beta<1,$ $\alpha+\beta\rightarrow 0.$ The holding time of any state is longer when the channel is negatively correlated and hence, the channel remains in a bad state for a longer time hurting the decoding delay. On the other hand, the state holding times are shorter for a positively correlated channel and state transitions are more frequent leading to a shorter decoding delay.
\section{Numerical Results}
\label{sec:num}
In this section, we provide numerical results to substantiate our analysis of the RLNC scheme, both for the case of the time invariant channel model and the time correlated channel model. As a representative setup for the case of the time invariant channel model, we let the OFF probability of erasure channels $p$ to be $0.1.$ Note that the broadcast capacity for this choice of $p$ is $(1-p)=0.9.$ For the time correlated channel model, we choose $\alpha=\beta=0.3.$ The broadcast capacity for this choice is $0.5.$ We note that the scaling behavior of the throughput and decoding-delay do not change
for any other choice of channel parameters. Our
numerical results are presented under two different
scenarios, the first focuses on confirming the phase
transition of the throughput scaling, and the second
focuses on substantiating the accuracy of our approximation for the mean decoding delay computed using extreme value theory.

\begin{figure}[htp]
\begin{minipage}{3in}
\centering
\includegraphics[height=2.5in]{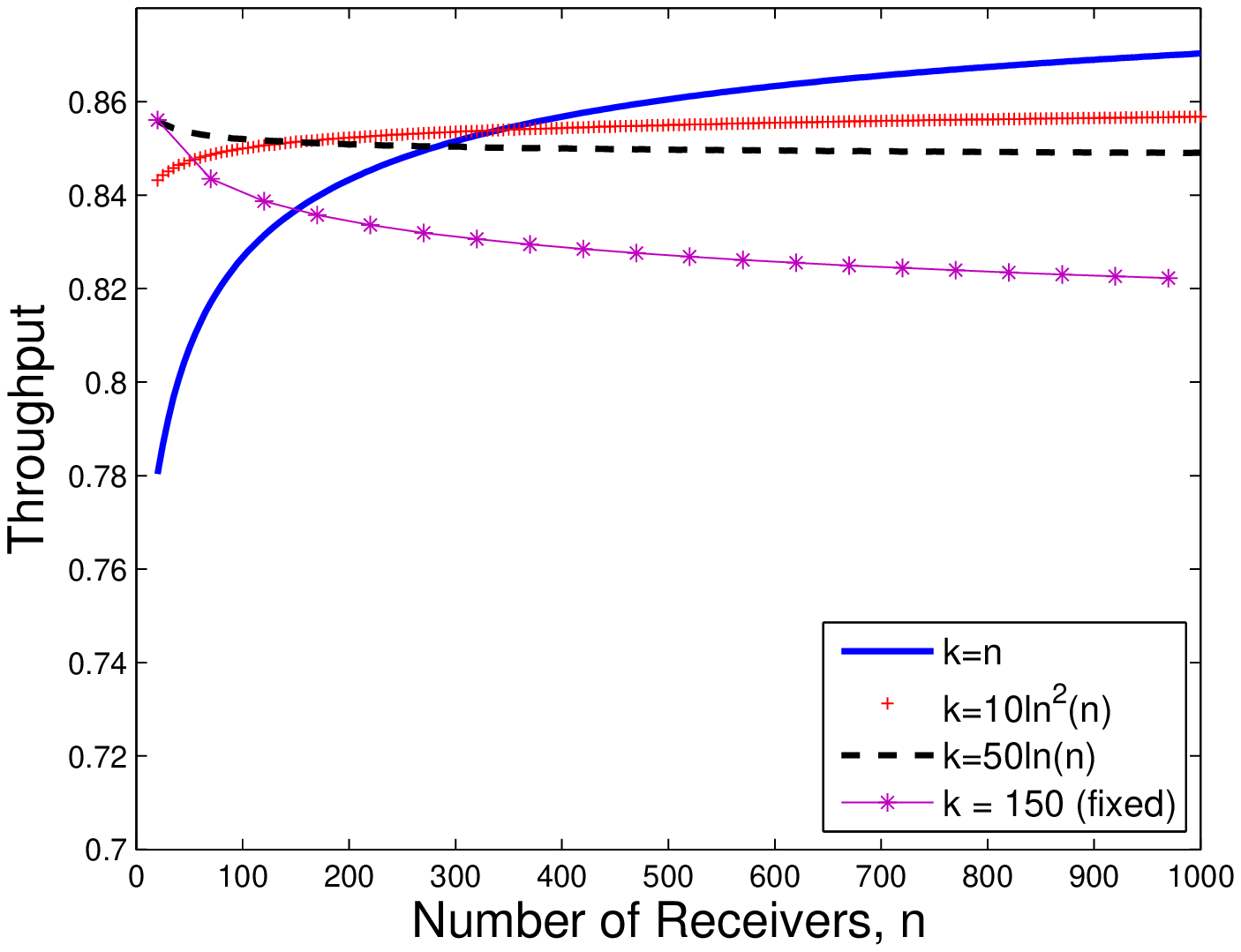}
\caption{Throughput behavior under different scalings of
$k$ with $n$ when $p=0.1$} \label{fig:gputall}
\end{minipage}
\hfill
\begin{minipage}{3in}
\centering
\includegraphics[height=2.5in]{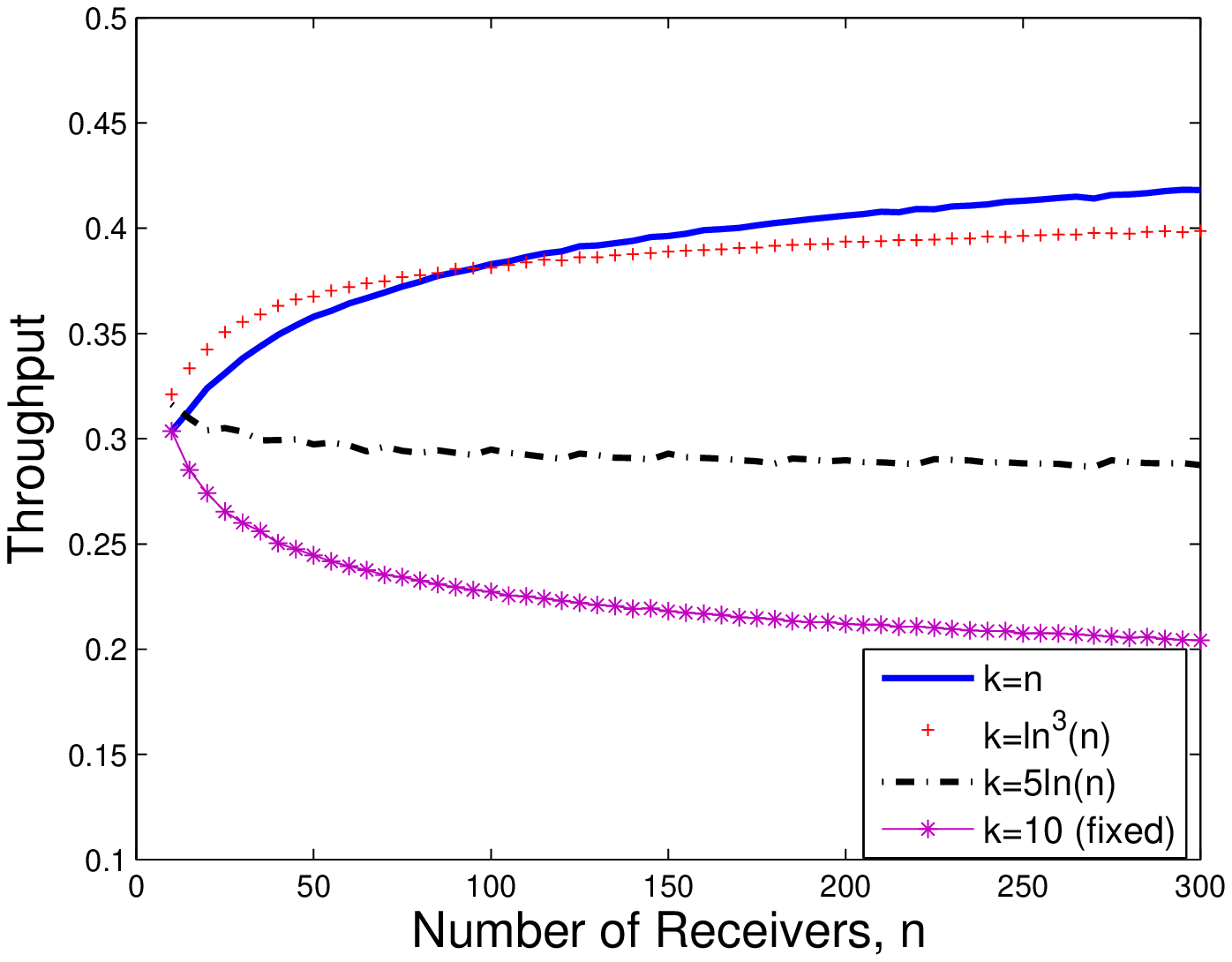}
\caption{Throughput behavior under different scalings of
$k$ with $n$ when $\alpha=\beta=0.3$ and capacity is $0.5$.} \label{fig:gputall2}
\end{minipage}
\end{figure}
\noindent{\it Study 1) Phase transition:} In this
study, we explore the phase transition law that is
suggested by Theorem~\ref{thm:main}. To that end,
Figure~\ref{fig:gputall} depicts the (broadcast)
throughput of RLNC in the actual system operation with
increasing $n$ for different types of scaling of $k$ for the time invariant erasure channel. We
see that this result is in perfect agreement with the phase
transition law: when $k=150$ and therefore scales slower
than $\ln(n),$ we see that the throughput decays towards
zero; when $k=50\ln(n),$ i.e. $k=\Theta(\ln(n)),$ the
throughput converges to a constant level as suggested by
Theorem~\ref{thm:main}; when $k=10 \ln^2(n)$ or $n,$ i.e.
$k=\omega(\ln(n)),$ the throughput increases toward the
broadcast capacity.

Figure~\ref{fig:gputall2} depicts the (broadcast)
throughput of RLNC in the actual system operation for the time correlated erasure channel model. Once again, we see that this result is in agreement with Theorem~\ref{thm:main}.\\
These two results also reveal that
the convergence rate of the performance to the capacity may be
increased by selecting a faster scaling of $k$ with respect
to $n$. Thus, Study~$1$ confirms the phase transition law suggested by the
our analysis. The next study is aimed at studying the
accuracy of our approximation for the mean decoding delay computed using extreme value theory.

\noindent{\it Study 2) Approximate Mean Decoding Delay:} Here, we consider two different scalings of $k$ with respect to $n$, and
compare the mean decoding-delay of the actual system to the approximate expression obtained using extreme value theory. In particular, we study the cases when $k=50\ln(n)$ and $k=n.$\\
\begin{figure}[htp]
\begin{minipage}{3in}
\centering
\includegraphics[height=2.5in]{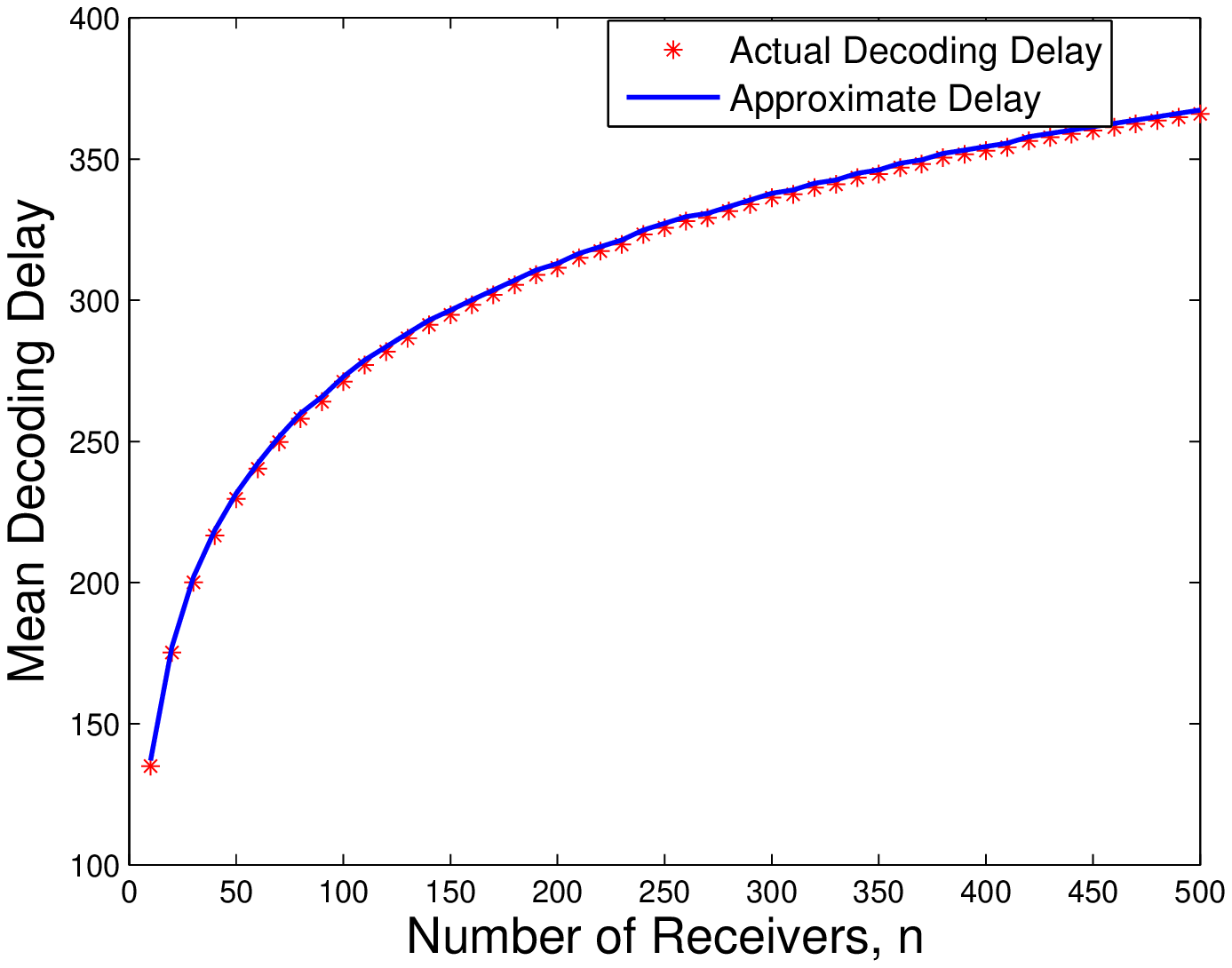}
\caption{Comparing Actual Mean Decoding Delay to Approximate
Mean for $p=0.1$ and $k=50\ln(n).$}
\label{fig:ctimelnN}
\end{minipage}
\hfill
\begin{minipage}{3in}
\centering
\includegraphics[height=2.5in]{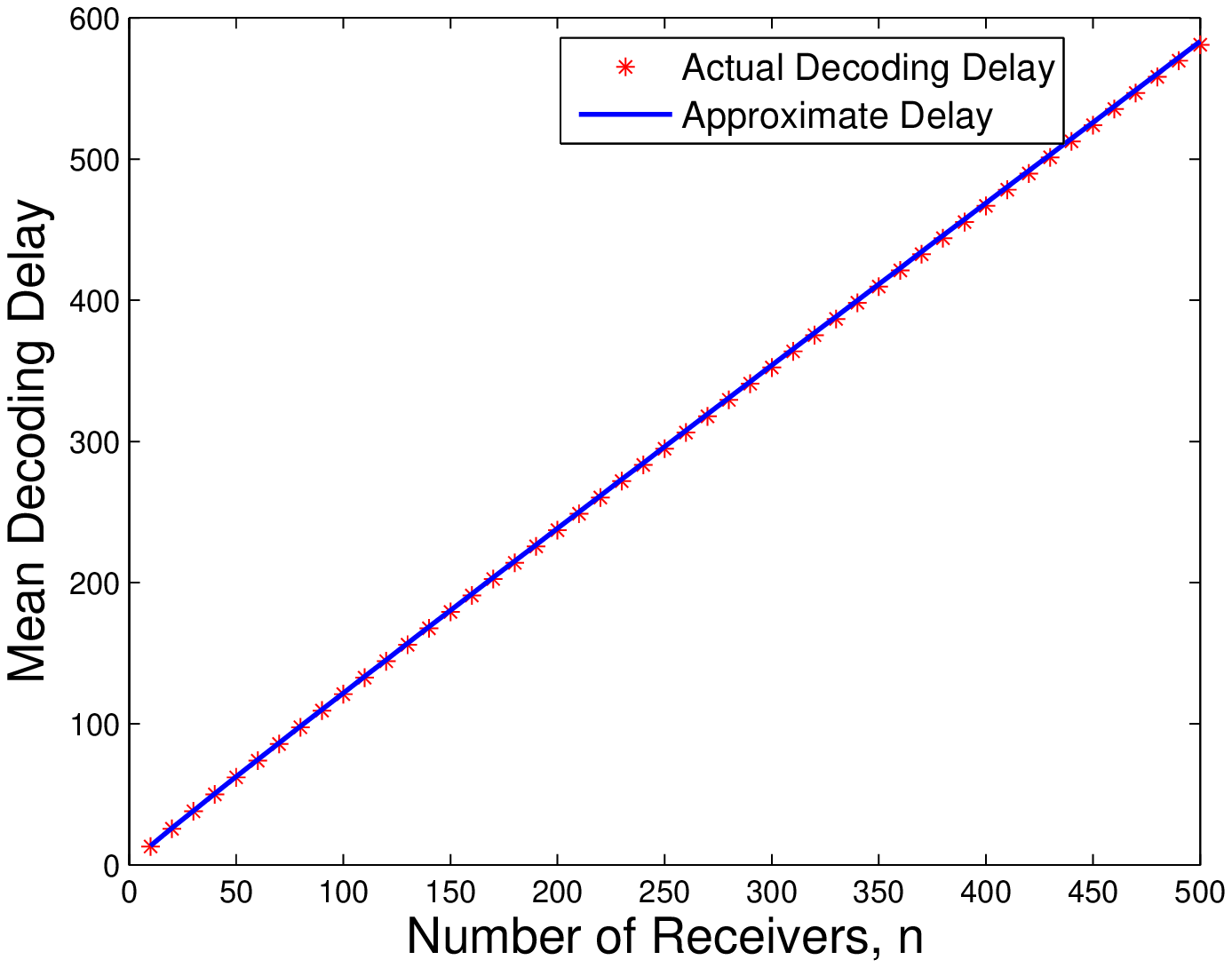}
\caption{Comparing Actual Mean Decoding Delay to Approximate Mean
for $p=0.1$ and $k=n.$}
\label{fig:ctimekn}
\end{minipage}
\end{figure}
First, we consider the case of time invariant channel model.
Figure~\ref{fig:ctimelnN} depicts mean decoding-delay performance when $k=50\ln(n)$ of the actual system behavior together with the approximate mean obtained using extreme value theory. This demonstrates the accuracy of our approximation even for small values of $n.$ We also see that a
throughput of approximately $0.85$ (see Figure~\ref{fig:gputall}) is achievable with this
scaling, leading to a decoding delay that scales only
logarithmically with the network size.\\
%
In comparison, Figure~\ref{fig:ctimekn} depicts the mean decoding-delay of the actual system and approximate mean when $k=n.$ Again, we observe
that our approximation is accurate and applicable to the actual
system performance, as predicted. In this fast scaling
scenario, we also observe that the throughput increases towards the capacity of $0.9$ (see Figure~\ref{fig:gputall}) instead of converging to a
constant level as in the case of scaling $k$ as $\ln(n)$. Yet, this
asymptotic optimality occurs at the cost of linearly
increasing decoding-delay performance.\\
\begin{figure}[htp]
\begin{minipage}{3in}
\centering
\includegraphics[height=2.5in]{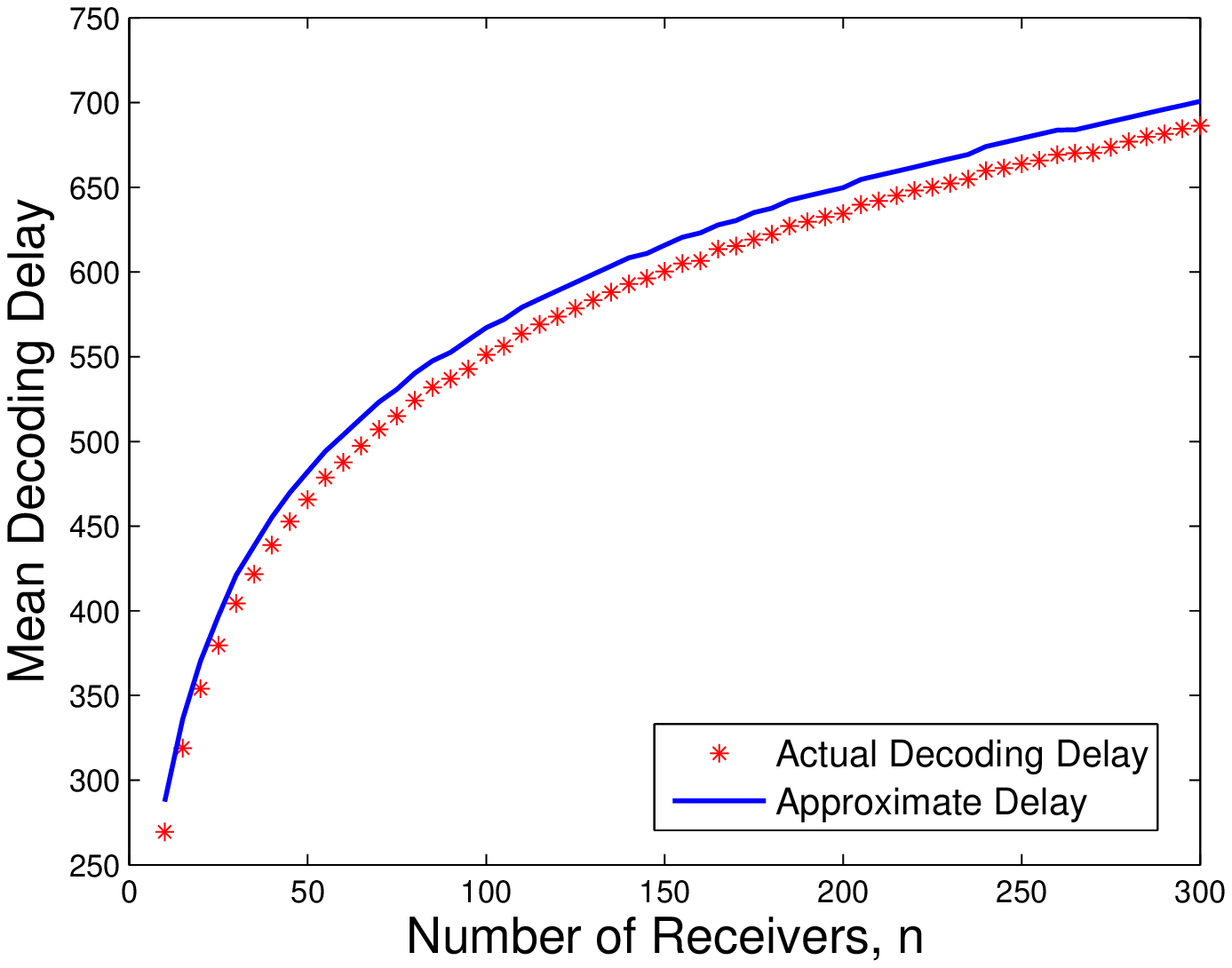}
\caption{Comparing Actual Mean Decoding Delay to Approximate
Mean for $\alpha=\beta=0.3$ and $k=50\ln(n).$}
\label{fig:ctimelnNcorr}
\end{minipage}
\hfill
\begin{minipage}{3in}
\centering
\includegraphics[height=2.5in]{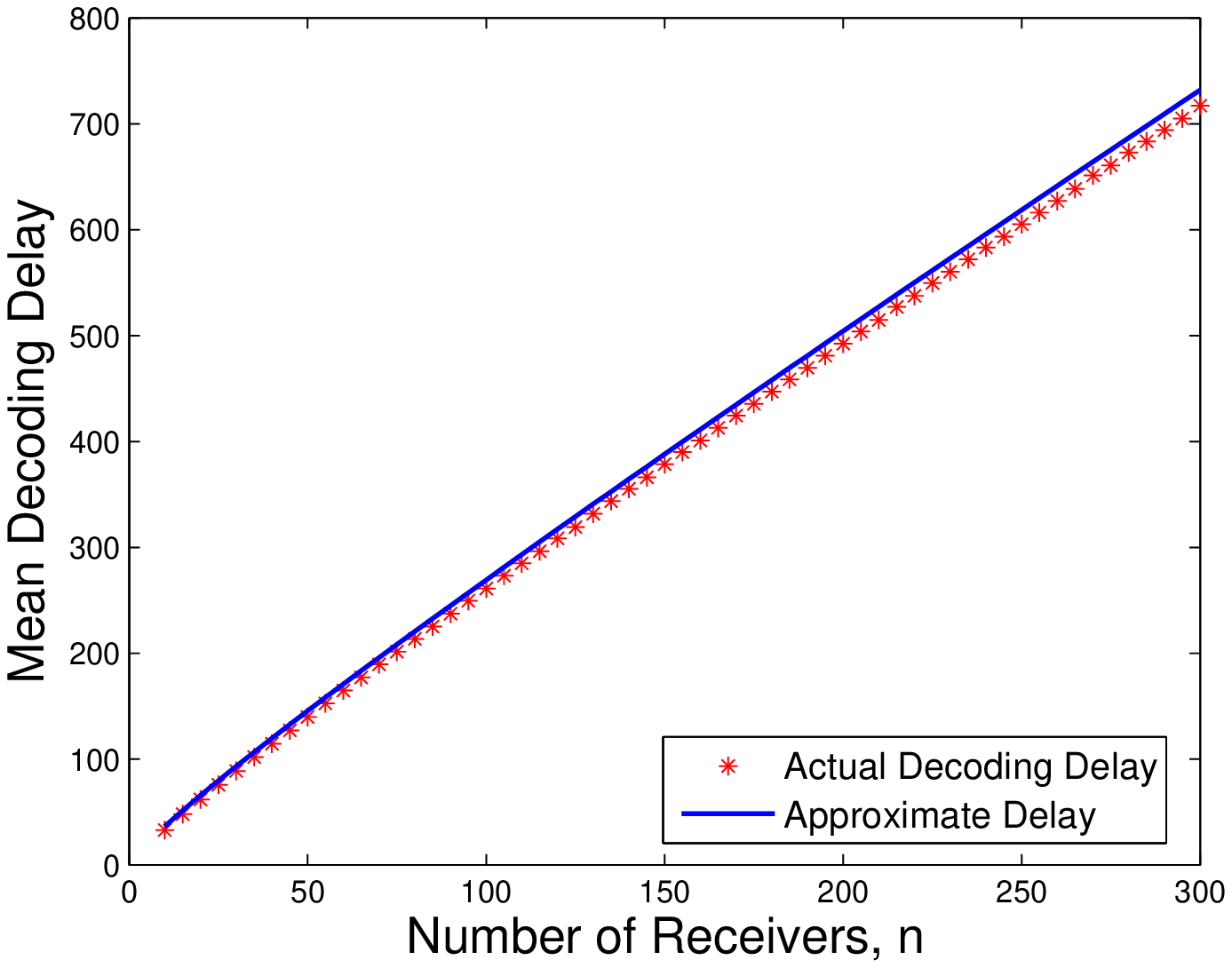}
\caption{Comparing Actual Mean Decoding Delay to Approximate Mean
for $\alpha=\beta=0.3$ and $k=n.$}
\label{fig:ctimekncorr}
\end{minipage}
\end{figure}
Next, we consider the case of time correlated channel model and compare the actual decoding delay of the RLNC scheme to the approximate expression in (\ref{eq:approxcorrmean}). Recall that the approximate decoding delay is derived under the assumption that the channel state of the system prior to the beginning of the current block transmission is $s=1.$ Nonetheless, we see, from Figures~\ref{fig:ctimelnNcorr} and \ref{fig:ctimekncorr}, that the approximation in (\ref{eq:approxcorrmean}) is an accurate characterization of the mean decoding delay of our system.\\
\begin{figure}[ht]
\centering
\includegraphics[height=2.5in]{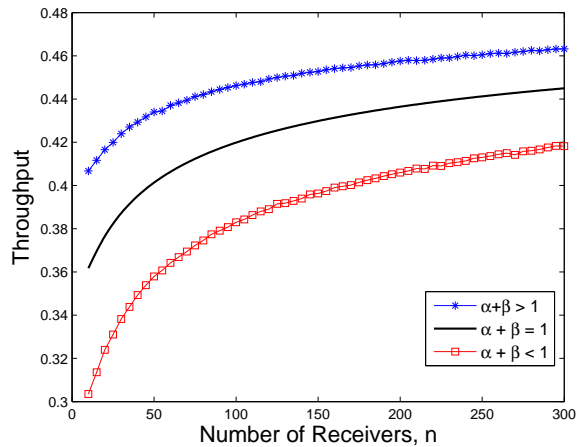}
\caption{Comparing the Rate of Convergence of Throughput when $k=n$ for three cases of Correlated Channels: $1)$ $\alpha=\beta=0.7,$ $2)$ $\alpha=\beta=0.5,$ and $3)$ $\alpha=\beta=0.3.$} \label{fig:comparen}
\end{figure}
Next, we compare the broadcast throughput performance of RLNC for positively and negatively correlated channels with the same broadcast capacity of $0.5.$ From Figure \ref{fig:comparen}, we see that the throughput performance is better for the case of positively correlated channels ($\alpha=\beta=0.7$) while it is worse for the case of negatively correlated channels ($\alpha=\beta=0.3$).
Overall, these numerical studies collectively confirm the accuracy of
estimating the mean decoding delay of RLNC using extreme value theory when $k$ scales as function of $n.$
\section{Extensions}
\label{sec:disc}
In this section, we discuss three important extensions of our analysis. First, we discuss how our analysis can be extended to LT codes. Next, we comment on the case where the channels are not independent across the users. Finally, we comment on the case of asymmetric channels where the channel of user $i$ has a steady state erasure probability of $p_i.$
\subsection{LT Codes}
LT codes~\cite{luby} are rateless codes designed such that receiving any $\nu_k(\delta_k)=k+O(\sqrt{k}\ln^2(k/\delta_k))$ encoded packets guarantees that the receiver can decode the original $k$ packets with probability $(1-\delta_k),$ where $\delta_k\in(0,1).$ Note that by decreasing $\delta_k,$ and hence increasing $\nu_k(\delta_k),$ we can increase the probability of successfully decoding at the receiver. Assume that, under the coding strategy of LT codes, the source starts transmitting the next block only after all the receivers receive $\nu_k(\delta_k)$ encoded packets each of the current block. The analysis of Section~\ref{sec:corranalysis} can be repeated for the case of LT codes by replacing $k$ with  $\nu_k(\delta_k)$ in the derivation of the bounds on the inter-renewal interval $\mathbb{E}[W]$ and the computation of mean and variance of the decoding delay. However, in the case of LT codes, the users that receive more than $\nu_k(\delta_k)$ encoded packets have a higher probability of decoding than the other users. Therefore, we have to modify our definition of throughput as follows,
\begin{definition}[(Broadcast) Throughput]
We let $R_i[t]$ denote the number of packets successfully decoded by the user $i$ in a total
of $t$ slots. Then, the (broadcast) throughput for a given
$n$ and $k$ obtained using the LT coding scheme, denoted as $\eta(n,k)$, is the
long-term average number of successfully transferred data packets to
all $n$ users. Hence, we have
\begin{equation}
\eta(n,k) = \lim_{t \rightarrow \infty} \frac{\sum_{i=1}^nR_i[t]}{nt}
\end{equation}
\end{definition}
The expected reward obtained by the source in one inter-renewal interval of the renewal process defined in Section~\ref{sec:corranalysisA} is given by $\displaystyle \frac{\sum_{i=1}^n\mathbb{E}[R_i]}{n}$ where $\mathbb{E}[R_i]$ is reward obtained due to user $i.$ Since each user can decode the current block with at least a probability of $(1-\delta_k),$ we have that, $\mathbb{E}[R_i]\ge k\mathbb{E}[M](1-\delta_k).$ Therefore, we have that, $\displaystyle \eta(n,k)\ge \frac{k\mathbb{E}[M](1-\delta_k)}{\mathbb{E}[W]}.$
Also, since the maximum reward that the source can obtain from each block transmission is $k,$ we have that $\displaystyle \eta(n,k) \le \frac{k\mathbb{E}[M]}{\mathbb{E}[W]}.$ If $\delta_k$ is such that $\ln(k/\delta_k)=o(k^{(1/4)}),$ then $ \displaystyle \lim_{k\rightarrow\infty} \frac{\nu_k(\delta_k)}{k}=1.$ We can now show the following result analogous to Theorem~\ref{thm:main}:
\begin{theorem}
Using the LT coding scheme, the broadcast throughput $\eta(n,k)$ of our system, when $\alpha,\beta >0,$ and $\alpha+\beta\ne 2,$ can be characterized as follows,

\begin{enumerate}
\item If $k=o(\ln(n)),$ then
\begin{equation}
\lim_{n\rightarrow \infty} \eta(n,k)= 0.
\end{equation}
\item If $k=\omega(\ln(n)),$ and if
\begin{enumerate}
\item $\delta_k=\delta$ for all $k,$ where $\delta\in(0,1),$ then
\begin{equation}
\liminf_{n\rightarrow \infty} \eta(n,k)\ge (1-p)(1-\delta).
\end{equation}
\item $\ln(k/\delta_k)=o(k^{(1/4)}),$ and $\displaystyle \lim_{k\rightarrow \infty} \delta_k=0,$ then
\begin{equation}
\lim_{n\rightarrow \infty} \eta(n,k)= (1-p).
\end{equation}
\end{enumerate}
\item Furthermore, if $k=\Theta(\ln(n)),$ and if
\begin{enumerate}
\item $\delta_k=\delta$ for all $k,$ where $\delta\in(0,1),$ then
\begin{equation}
\liminf_{n\rightarrow \infty} \eta(n,k)\ge r(1-p)(1-\delta),
\end{equation}
\item $\ln(k/\delta_k)=o(k^{(1/4)}),$ and $\displaystyle \lim_{k\rightarrow \infty} \delta_k=0,$ then
\begin{equation}
\liminf_{n\rightarrow \infty} \eta(n,k)\ge r(1-p),
\end{equation}
\end{enumerate}
for some $0<r<1.$\\
\end{enumerate}
\end{theorem}
Recall that $(1-p)$ is an upper bound on the throughput achievable by any scheme. A choice of $\delta_k$ that satisfies the conditions $1)$ $\ln(k/\delta_k)=o(k^{(1/4)})$ and $2)$ $\displaystyle \lim_{k\rightarrow \infty} \delta_k=0$ is given by $\delta_k=\frac{1}{\ln(k)}.$ Hence, under the scaling regime of $k=\omega(\ln(n)),$ this choice of $\delta_k$ would ensure that the throughput approaches $1-p$ as $n\rightarrow\infty.$
\subsection{Dependent channel model}
Next, consider the case of symmetric time-invariant channels that are dependent across the users. For all $i,$ let $C_i[t]=0$ with a probability $p$ and $C_i[t]=1$ with probability $(1-p).$ Lemma~\ref{lem:main} holds true even when the random variables $\chi_i,$ $i=1,\ldots,n$ are dependent. However, Lemmas~\ref{lem:singleub},~\ref{lem:corrub1} and~\ref{lem:corrlb} are not valid anymore. Nonetheless, we can show that if $k=\omega(\ln(n)),$ then $\displaystyle \lim_{n\rightarrow \infty} \eta(n,k)=1-p$ even when the user channels are dependent. Next, we prove this sufficient condition to achieve broadcast capacity.

Once again, we drop the superscript $(j)$ in the following discussion. Let $\mu(k)=\frac{k}{1-p}$ and $\sigma^2(k)=k\frac{p}{(1-p)^2}$. Define
$\tilde{Y_i}(k)=\frac{Y_i(k)-\mu(k)}{\sigma(k)}$.
Note that $U$ can be rewritten as follows,
\begin{equation}
\label{eqn:delaynew}
U= \mu(k) + \sigma(k)\max_{1\le i \le n}\tilde{Y_i}(k)
\end{equation}
Consider the case when $k= f(n)\ln(n)$ for some function $f(n)> 0$. Note that in our analysis we treat $k$ as a continuous function of $n$. This assumption does not seriously affect the analysis and can easily be relaxed. Choose $\tau=b\sqrt{\ln(n)}$ where $b$ is a constant such that $0 < \tau\le\sigma(k)\ln(\frac{1}{1-p})$. Such a $\tau$ lies within the radius of convergence of the m.g.f of $\tilde{Y_1}(k),$ which is then given by:
\begin{align}
\mathbb{E}[e^{\tau\tilde{Y_1}(k)}] &= \exp\left(\frac{\tau k}{\sigma(k)}\right)\exp\left(\frac{-\tau\mu(k)}{\sigma(k)}\right)\left[\frac{1-p}{1-p\exp\left(\frac{\tau}{\sigma(k)}\right)}\right]^k \nonumber\\
\label{eqn:mgf}
&= \left[\frac{(1-p)\exp\left(\frac{-b\sqrt{p}}{\sqrt{f(n)}}\right)}{1-p\exp\left(\frac{b(1-p)}{\sqrt{pf(n)}}\right)}\right]^{f(n)\ln(n)}
\end{align}
Now, using (\ref{eqn:ineq1}) with $\tau=b\sqrt{\ln(n)}$, and, (\ref{eqn:delaynew}), and (\ref{eqn:mgf}), we can bound the expected decoding delay and, hence, the throughput as follows:
\begin{align}
\mathbb{E}[U]&\le \frac{g(n)\ln(n)}{1-p},\\
\label{eq:gputfinal}
\eta(n,k)&\ge h(n)(1-p),
\end{align}
where (\ref{eq:gputfinal}) follows from (\ref{eq:gput}), and $g(n)$ and $h(n)$ are given as
\begin{align*}
g(n)&=f(n)+\frac{\sqrt{f(n)p}}{b}\left(1+f(n)\ln\left(\frac{(1-p)\exp\left(\frac{-b\sqrt{p}}{\sqrt{f(n)}}\right)}{1-p\exp\left(\frac{b(1-p)}{\sqrt{pf(n)}}\right)}\right)\right),\\
h(n)&=\frac{f(n)}{g(n)}.
\end{align*}
First, let $f(n)=\hat{b}$ where $\hat{b}\ge 0$ is a constant independent of $n$. Then $g(n)$ and $h(n)$ are constants independent of $n$ with $h(n)<1$ ($h(n)$ can be expressed explicitly in terms of $b,$ $\hat{b}$ and $p$). From (\ref{eq:gputfinal}), we see that the scaling regime of $k=\Theta(\ln(n))$ guarantees that $\displaystyle \liminf_{n\rightarrow \infty}\eta(n,k)$ is a non-vanishing fraction of the broadcast capacity $(1-p)$. Next, let $f(n)$ be such that $f(n)\rightarrow \infty$ as $n\rightarrow \infty.$ Then,
\begin{align*}
\lim_{n\rightarrow \infty} \frac{g(n)}{f(n)}-1 &= \lim_{n\rightarrow \infty} \frac{\sqrt{p}}{b\sqrt{f(n)}}\left(1+f(n)\ln\left(\frac{(1-p)\exp\left(\frac{-b\sqrt{p}}{\sqrt{f(n)}}\right)}{1-p\exp\left(\frac{b(1-p)}{\sqrt{pf(n)}}\right)}\right)\right)\nonumber\\
&= \lim_{n\rightarrow \infty} \frac{\sqrt{p}}{b}\sqrt{f(n)}\ln\left(\frac{(1-p)\exp\left(\frac{-b\sqrt{p}}{\sqrt{f(n)}}\right)}{1-p\exp\left(\frac{b(1-p)}{\sqrt{pf(n)}}\right)}\right)\nonumber\\
&= \lim_{n\rightarrow \infty} \frac{-\sqrt{p}}{b}\sqrt{f(n)}\left(1-\frac{(1-p)\exp\left(\frac{-b\sqrt{p}}{\sqrt{f(n)}}\right)}{1-p\exp\left(\frac{b(1-p)}{\sqrt{pf(n)}}\right)}\right)\nonumber\\
&= \frac{-\sqrt{p}}{b(1-p)}\lim_{n\rightarrow \infty} \sqrt{f(n)}\left(1-p\exp\left(\frac{b(1-p)}{\sqrt{pf(n)}}\right)-(1-p)\exp\left(\frac{-b\sqrt{p}}{\sqrt{f(n)}}\right)\right)\nonumber\\
&=\frac{-\sqrt{p}}{b(1-p)}\lim_{n\rightarrow \infty} \left(-\frac{b\sqrt{p}(1-p)}{\sqrt{f(n)}}\exp\left(\frac{b(1-p)}{\sqrt{pf(n)}}\right)-\frac{-b\sqrt{p}(1-p)}{\sqrt{f(n)}}\exp\left(\frac{-b\sqrt{p}}{\sqrt{f(n)}}\right)\right)\nonumber \tag{iii}\\
\label{eq:limits}
&= 0,
\end{align*}
where (iii) is obtained using $L'H\hat{o}spital's$ rule. \\
Hence, we have $h(n)\rightarrow 1$ and $\eta(n,k)\rightarrow (1-p)$ as $n\rightarrow \infty$ proving it is sufficient that $k=\omega(\ln(n))$ to achieve the broadcast capacity $(1-p).$\\
We can easily see that it is no longer true that if $k=o(\ln(n)),$ then $\displaystyle \lim_{n\rightarrow \infty} \eta(n,k)=0.$ For example, consider the trivial case of perfectly correlated user channels. Then the system reduces to the case where there is a single user channel and hence, the throughput, $\eta(n,k),$ of this system does not depend on $n.$ Also, the sufficient condition of $k=\omega(\ln(n))$ is unnecessary in this trivial case as it is always true that $\eta(n,k)\rightarrow(1-p)$ as $k\rightarrow\infty.$  Although the sufficiency condition appears to be a rather weak condition for the trivial example of perfectly correlated channels, it is still useful when the system has weak correlations.
\subsection{Asymmetric channel model}
Consider the case of asymmetric time-invariant channels where $C_i[t]=0$ with a probability $p_i$ and $C_i[t]=1$ with probability $(1-p_i).$ The channels of all users are assumed to be independent of each other. Suppose that $\displaystyle p_0=\sup_i p_i$ exists. Then the broadcast capacity of our system is $(1-p_0).$ Let $\mathbb{E}[U_1]$ and $\eta_1(n,k)$ denote the mean decoding delay and throughput of this system. We can compare this system with the symmetric system with erasure probabilities $p_0.$ Let $\mathbb{E}[U_0]$ and $\eta_0(n,k)$ denote the mean decoding delay and throughput of the symmetric system. By comparing the cumulative distributions of the decoding delay of the two systems, we can show that $\mathbb{E}[U_1]\le\mathbb{E}[U_0].$ Consequently, $\eta_0(n,k)\le\eta_1(n,k)\le 1-p_0.$ From Theorem~\ref{thm:main}, we have that, if $k=\omega(\ln(n)),$ then $\displaystyle \lim_{n\rightarrow\infty}\eta_0(n,k) = 1-p_0$ and hence, $\displaystyle \lim_{n\rightarrow\infty}\eta_1(n,k) = 1-p_0.$ \\
The case of asymmetric time-correlated channels is technically challenging since it is not easy to obtain a closed form expression for the mean decoding delay as in the case of time-invariant channels. 

\section{Conclusion}
\label{sec:conc} We have investigated the throughput and
decoding delay performance of RLNC in a wireless broadcast
setting as the coding window size $k$ scales with the number of receivers $n$ for a time correlated erasure channel model. We
noted that the broadcast throughput of RLNC vanishes for
any fixed $k$ as the system size increases. Hence, it is
important to understand the scaling of $k$ as a function of
$n$ that will guarantee a non-vanishing throughput.

Our analysis revealed a phase transition in the performance
of our system, namely, if $k$ increases slower
than $\ln(n)$, the throughput goes to zero as $n$
increases. However, on increasing $k$ faster than $\ln(n)$,
the throughput approaches the maximum achievable broadcast
throughput of $(1-p)$. Also, $k=\Theta(\ln(n))$ ensures a
constant fraction of the maximum achievable broadcast
throughput for the our system. Further, we have provided approximate expressions for the mean decoding delay under the scaling regime of $k=\omega(\ln(n))$ using extreme value theory. We have shown through numerical results that our approximation is accurate even for small values of $n.$  We have also shown that our analysis can be extended to other rateless block coding schemes such as the LT codes. In particular, by choosing $\delta_k$ such that $\displaystyle \lim_{k\rightarrow \infty} \delta_k=0,$ we see that under the scaling regime of $k=\omega(\ln(n)),$ LT codes achieve the broadcast throughput of $(1-p)$ as $n\rightarrow\infty.$ Further, we commented on the extension of our analysis for the cases of dependent channels across users, and asymmetric channel model.

\bibliographystyle{IEEEtran}
\bibliography{IEEEabrv,myrefs}
\end{document}